\documentstyle[proceedings,psfig]{crckapb} 

\def\Lbsun{\hbox{$\rm\thinspace L_{B\odot}$}}
\def\Mbsun{\hbox{$\rm\thinspace M_{B\odot}$}}

\begin{opening}
\title{
The Type Ia Supernova Rate at $\lowercase{ z} \sim 0.4$}
\subtitle{The Supernova Cosmology Project: IV}

\author{R. Pain}
\author{$^{1,2}$}
\author{I. M. Hook$^1$}
\author{S. Perlmutter}
\author{S. Deustua}
\author{S. Gabi}
\author{G. Goldhaber}
\author{D. Groom}
\author{A. G. Kim}
\author{M. Y. Kim}
\author{J. C. Lee}
\author{C. R. Pennypacker}
\author{I. A. Small}
\institute{E. O. Lawrence Berkeley National Laboratory \& Center
for Particle Astrophysics,
University of California, Berkeley}
\author{A. Goobar}
\institute{University of Stockholm}
\author{R. Ellis}
\author{R. McMahon}
\author{K. Glazebrook$^3$}
\institute{Institute of Astronomy, Cambridge University}
\author{B. Boyle}
\author{P. Bunclark}
\author{D. Carter}
\author{M. Irwin}
\institute{Royal Greenwich Observatory}

\end{opening}

\runningtitle{The Type Ia Supernova Rate at $\lowercase{z}\sim 0.4$}

\begin{document}

\footnotetext[1]{Presented by 
I. Hook ({\it imh@mh1.lbl.gov}) and 
R. Pain ({\it rpain@lpnax1.in2p3.fr})}
\footnotetext[2]{Current address:  CNRS-IN2P3, University of Paris}
\footnotetext[3]{Current address: Anglo-Australian Observatory}

\begin{abstract} We present the first measurement of the rate of Type
Ia supernovae at high redshift. The result is derived using a large
subset of data from the Supernova Cosmology Project
as described in more detail at this meeting by Perlmutter et al. (1996).
We present our methods for
estimating the numbers of galaxies and the number of solar
luminosities to which the survey is sensitive, the supernova detection
efficiency and hence the control time. We derive a rest-frame Type Ia
supernova rate at $z\sim0.4$ of $0.82\ {^{+0.54}_{-0.37}}\
{^{+0.42}_{-0.32}}$ $h^2$ SNu where the first uncertainty is 
statistical and the second includes systematic effects.
\end{abstract}
\section{Introduction}
Beginning with the discovery of SN 1992bi (Perlmutter et al 1995), we
have developed search techniques and rapid analysis methods that allow
systematic discovery and follow up of ``batches'' of high-redshift
supernovae.
The observing strategy developed compares large numbers
of galaxies in each of $\sim$50 fields observed twice with a
separation of $\sim$3 weeks.
This search schedule makes it possible to precisely
calculate the ``control time,'' the effective time during which the
survey is sensitive to a Type Ia event.

For this analysis, we have studied a set of 52 similar search fields
observed in December 1993 and January 1994. 
The data were obtained using the ``thick'' $1242\times 1152$ EEV5
camera at the 2.5m Isaac Newton Telescope, La Palma.  The projected
pixel size is $0.56''$, giving an image area of approximately
$11'\times 11'$. Exposure times were 600s in the $R_{Mould}$ filter,
and the images typically reach a $3\sigma$ limit of $R=23$mag. Seeing
was typically around $1.4''$. 
Many of the fields were
selected due to the presence of a high-redshift cluster ($z\sim 0.4$).
Suitable clusters and their redshifts were taken from Gunn, Hoessel \&
Oke (1986) and from the ROSAT catalog. 
The total useful area covered in this study is 1.73 sq deg.

The analysis procedure and method for finding SNe can be summarized as
follows.  For most fields, two first-look ``reference'' images were
obtained and for all fields two second look ``search'' images were
obtained $2-3$ weeks after the reference images.  The images were
flat-fielded and zero-points for the images were estimated by
comparison with E (red) magnitudes of stars in the APM (Automated
plate measuring facility, Cambridge, UK) POSSI catalog (McMahon \&
Irwin 1992).  The search images were combined (after convolution to
match the seeing of the worst of the four images) and the combined
reference images were subtracted from this after scaling in intensity.
The resulting difference image for each field was searched for SN
candidates.

In this subset of the search data three SNe were found with redshifts 
0.354, 0.375 and 0.420 determined from spectra of the
host galaxies. For the purposes of this analysis, we assume these are
all Type Ia SNe (see Perlmutter et al. 1996).
The method used to calculate the rate can be divided into two main
parts: (i) estimation of the number of galaxies and the
total stellar luminosity (measured in $10^{10}$\Lbsun) to which the
survey is sensitive and 
(ii) estimation of the SN detection efficiency and hence the control time.

\section{Galaxy counts}
\begin{figure} 
\psfig{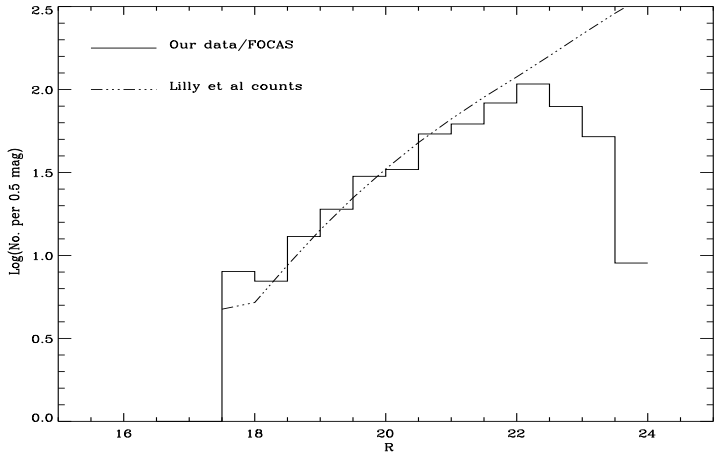}
\caption{Number of galaxies as a function of magnitude determined from
one of our non-cluster images using FOCAS. The dashed-dot line shows
the counts derived from the analysis of Lilly et al (1995), integrated
over the redshift range $0<z<2$, and normalized to the image area of
0.03 sq deg.}
\label{nofm} 
\end{figure}

In order to compare the distant SN rate with local equivalents, we
need to know the redshifts of the galaxies we have surveyed. 
In this work we use the galaxy counts derived
from the analysis of Lilly et al (1995) to estimate the number of
galaxies sampled as a function of redshift.
$R$ band counts as a function of redshift were calculated by Lilly
based on the analysis of magnitude--redshift data obtained in the
Canada-France Redshift Survey (Lilly et al, 1995 and references
therein). To check that the assumed distribution of galaxies with $R$
magnitude and redshift, $N(z,R)$, give reasonable galaxy counts 
compared to our data, 
we have plotted the number of field galaxies classified by the
FOCAS software package, as a function of apparent magnitude, on one of
the search images that was {\em not} targeted at a galaxy cluster.
The R-band galaxy counts given by the analysis of Lilly et al (1995)
integrated over the redshift range
$0<z<2$ (dash-dotted line) are shown on the same scale, assuming an effective
area for this image of 0.03 sq deg (Figure~\ref{nofm}).

Many of our search fields were chosen specifically to target
high-redshift clusters.  For each of these fields, we estimate the
number of cluster galaxies by counting galaxies as a function of $R$
magnitude in a box of size $500\times 500$ pixels centered
approximately on the center of the cluster as estimated by eye from
the images. The counts in a similar box in a region of the image away
from the cluster were subtracted from the cluster counts to give the
cluster excess counts as a function of $R$ mag.
Typically a cluster
contributes 10\% of the galaxy counts on an image.  We assign these
galaxies to the cluster redshift, and add the cluster contribution to
the $N(z,R)$ for that image given by the models.

Values of reddening due to the Galaxy were supplied by D. Burstein
(derived from Burstein \& Heiles, 1982) for each field and applied to
the data assuming $R_V = 3.1$ and $A_R/A_V = 0.751$ (Cardelli et al,
1989).

\section{Control times and detection efficiencies}
To calculate the control times we assumed that SN magnitude as a function
of time follows the average of the best-fit, time-dilated and
$K$-corrected Leibundgut (1988) Type Ia template light curves (the generalized 
$K$ correction described by Kim et al. (1996) and Kim, Goobar \& 
Perlmutter (1995), was used here).
The control time is then given by the weighted sum
of days during which the SN can be detected, where the weighting 
is by the detection efficiency, $\epsilon$.
 
This detection efficiency is a function of the
magnitude difference $\Delta m$ (zero if $\Delta m$ is negative),
which is itself a function of time $t$ relative to maximum, and
$\delta t$, the time separation of the search and reference images.
It depends 
on the quality of the subtracted images (seeing, transmission)
together with the detailed technique (convolution, selection criteria)
used to extract the signal (SNe candidates) from the background
(cosmic rays, asteroids, bad subtractions, etc). In addition, there is
a slight dependence on the host galaxy magnitude.  The detection
efficiency was calculated using a Monte-Carlo method. A synthetic
image was created for every field by adding simulated supernovae to
the search images using point-spread-functions drawn from each search image. 
The reference images were subtracted from the
synthetic search images using exactly the same software as used for the
supernova search and the number of simulated
SNe that satisfy the selection criteria was determined. This technique
allows us to measure detection efficiencies as a function of
supernovae magnitude individually for every field, thus taking into
account the other parameters mentioned above.

\begin{figure}
\psfig{figure=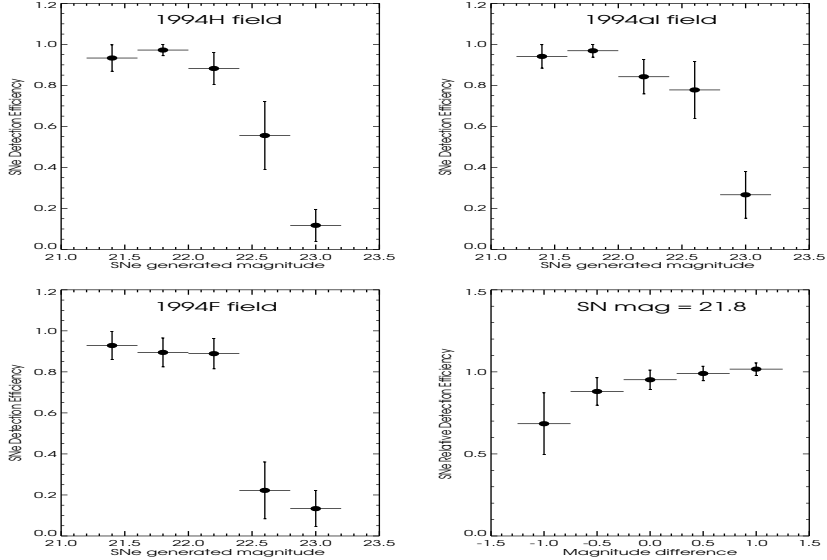,height=3.0in,width=4.5in}
\caption{(a)-(c) Detection efficiency as a function of relative SN
to host surface brightness. (d) Detection efficiency as a function of
magnitude for the three difference images in which SNe were found. The
vertical error bars show the $1\sigma$ statistical uncertainty, and
the horizontal bars show the bin ranges.}  \label{eff}
\end{figure}

Figure~\ref{eff}(a)-(c) shows the fractional number of simulated SN
recovered, as a function of SN magnitude (at detection) for the three
fields in which SNe were found. Figure~\ref{eff}(d) shows the
efficiency as a function of relative surface brightnesses of the SN
and host galaxy.  This last parameter gives an indication of the
effect of SN location with respect to the host galaxy. Although this
is a small effect, it was taken into account. For a typical field the
detection efficiency is over 85\% for any added fake stellar object
brighter than $R=22.0$ magnitude (Note that the more recent searches of 
this project have worked with significantly deeper images).

\section{SN Ia Rates}
\subsection{Survey rate}
Before calculating the luminosity-weighted SN rate, we first determine 
a ``survey rate'' of SN
discoveries that a search for Ia SNe can expect to obtain, per square
degree. We give the rate in a range of 1 mag in $R$, centered on the
mean peak $R$ magnitude of the 3 SNe found in this search, $R=21.8$. The
survey rate is given by
 $$ survey\ rate\ (21.3<R<22.3) = {N_{SN}\over{\sum_i area_i \times
\Delta T_i}}.$$ where $N_{SN}=3$ is the number of SNe we found in the
1 magnitude range, and $\Delta T_i$ is the control time for field $i$,
computed for a SN with magnitude $R=21.8$ at maximum.  For example a
value of $\Delta T_i= 21$ days was found for the field containing
SN1994H observed on 1993 December 19 and 1994 January 12, 24 days
apart. We measure a survey rate for $21.3 < R < 22.3$ of $ 34.4\ {
^{+23.9}_{-16.2}}$ SNe\ $year^{-1} deg^{-2}$ (the error quoted is
statistical only). In practice this translates to 1.73 SNe per square
degree discovered with a 3 week baseline, in data with limiting
magnitude $R\sim 23$ ($S/N\sim3$ for $R$=23).

\subsection{Rate in SNu}
To calculate the rate in SNu, we compute
the expected redshift distribution of SNe, $N_{exp}(z)$, which is
proportional to the observed SN rate, $r_{SN}(1+z)^{-1}$, where
$r_{SN}$ is the rate in the rest-frame of the supernovae. It 
is given by $$N_{exp}(z)={r_{SN}\over{1+z}}\sum_R\sum_i
N_{gal}(z,R)_i \times L_B (z,R) \times \Delta T_i (z,R) $$ where $i$
runs over all fields, $R$ is the galaxy apparent $R$ magnitude and $L_B$
is the galaxy rest-frame B band luminosity in units of
$10^{10}$\Lbsun.  
The control times $\Delta T$, in units
of centuries, have been calculated for each field in bins of $z$ and
$R$ (the size of the bins is 0.5 mag in $R$, 0.05 in $z$). 

To compute the rest-frame B band galaxy luminosities from apparent $R$
magnitudes, we used $B-R$ colors and $B$-band $K$ corrections supplied by 
Gronwall \& Koo. Note that the combined color, $K$ and evolution correction
is small in the redshift range of interest ($0.3-0.5$) since the
observed $R$ band is close to the rest-frame B-band.  
In this calculation $\Mbsun=5.48$ and $q_0=0.5$ were assumed.
\begin{figure} 
\psfig{file=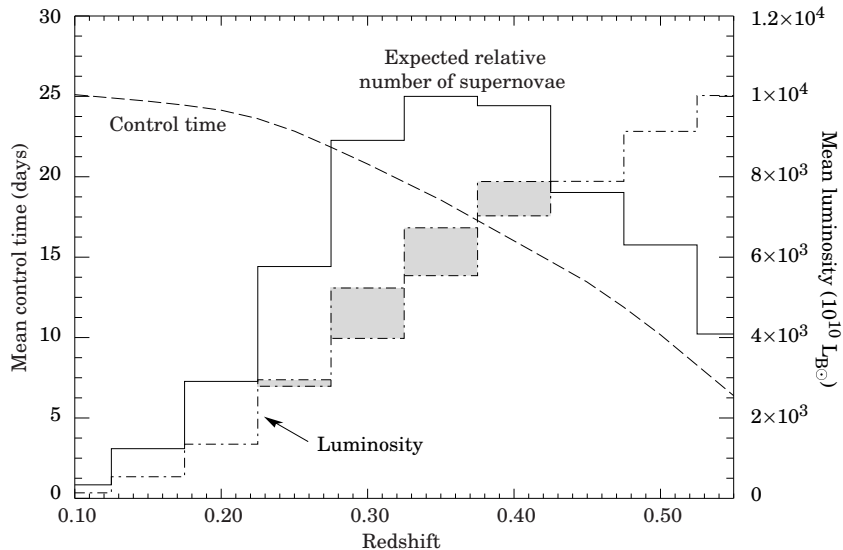,width=4.5in,height=3.0in}
\caption{The expected number of supernovae as a function of $z$ (solid
histogram) together with the overall control time (dashed curve) and
the luminosity-weighted number of galaxies (dash-dot histogram). The
contribution to the luminosity from clusters is shown by the shaded
area. The December 1993-January 1994 search data was most likely to
find SNe with redshifts between $z=0.3$ and $z=0.4$. Between $z=0.3$
and $z=0.55$, the search was more than 50\% efficient. Note that our
more recent searches go deeper than this data.}
\label{exp} 
\end{figure}
The rest frame supernovae rate $r_{SN}$ at $z\sim0.4$ was obtained
by fitting the redshift distribution of observed SNe to the expected
distribution, $N_{exp}(z)$ using a maximum-likelihood fit using
Poisson statistics.  The mean redshift corresponding to this rate is
$<z>=0.38$. We derive a value for the SN rate of
$$r_{SN}(z=0.38)=0.82\ ^{+0.54}_{-0.37}\ h^2\ \rm SNu.$$
where the error is statistical only.

\subsection{Systematic Uncertainties}

We have studied various sources of systematic uncertainty. Table 
~\ref{systab} summarizes their contributions. 
\begin{table} 
\caption{Systematic Uncertainties.
Uncertainties in the rate are in $h^2$ SNu.}
\begin{center}
\begin{tabular}{|lc|}
\hline
Source                                  & uncertainty \cr 
\hline
Luminosity estimate                     &   0.09 \cr
Cluster contribution                    &   0.02 \cr
Galaxy extinction                       &   0.01 \cr
APM calibration                         &   0.10 \cr
Detection efficiency                    &   0.08 \cr
non-standard SNe                        &   0.29 \cr
Scanning efficiency                     &$-0.00+0.27$ \cr
\hline
Total syst. uncertainty                 &$-0.32+0.42$ \cr
\hline
\end{tabular}
\end{center}
\label{systab}
\end{table}
For given assumed galaxy counts,
the main contribution comes from the fact that Type Ia SNe are not 
perfect standard candles. 

We also tested the sensitivity of our result to the assumed form of
galaxy evolution, $N(z,R)$, by recalculating the rate using the model
of Gronwall \& Koo (1995) and that used by Glazebrook et al (1995).
These give values for the rate of $1.61^{+1.05}_{-0.73} h^2$ SNu and
$1.27^{+0.83}_{-0.57} h^2$ SNu respectively.  Although there is a
large difference in the results derived using the counts of Lilly et
al compared with the other models, we choose not to include this in
the systematic uncertainty. For the purposes of this analysis we use
the Lilly et al counts, since these are based on data that are
well-matched to our survey in magnitude and redshift range, and only
small amount of extrapolation was required in converting from the $I$
to $R$ band.

Host galaxy inclination and extinction effects could not be estimated
following the analysis used for nearby searches because of the
different detection techniques involved. A correct estimation would
require modeling of galaxy opacities, which is beyond the scope of
this paper. We therefore compare our uncorrected value with
uncorrected values for nearby searches,
 
Altogether, we estimate the total systematic error to be $^{+0.32}_{-0.42}$ 
and we assume the Lilly et al counts for $N(z,R)$.

\section{Discussion and Conclusion}

This measurement is the first direct measurement of the Type Ia rate
at high redshift.  In their pioneering work searching for high
redshift supernovae, Hansen et al. (1989) discovered a probable Type
II event at $z = 0.28 $ and a Type Ia event at $z = 0.31 $ (N\o
rgaard-Nielsen et al., 1989), however no estimates of SN Ia rates were
published based on this discovery.

Nearby Supernovae rates have been carefully reanalyzed recently
(Cappellaro et al. 1993a \& 1993b, Turatto et al., 1994, Van den Bergh
and McClure, 1994, Muller et al. 1992) using more precise methods for
calculating the control times and correcting for inclination and
over-exposure of the nuclear regions of galaxies in photographic
searches.  The rate obtained for Type Ia SNe are now consistent among
these groups and vary between 0.2 $h^2$~SNu and 0.7 $h^2$~SNu
depending on the galaxy types (E, Sa etc., higher rates are found in
later type galaxies).  Taking into account the facts that at
$z\sim$0.4 the ratios of galaxy type are different and using the Type
Ia rates reported in Cappellaro et al (1993b), we calculate a local
rate of $0.53 \pm 0.25$ $h^2$ SNu for the mix of galaxies expected at
$z\sim0.4$.  Our measured value of $0.82 {^{+0.68}_{-0.49}}h^2$ SNu
(where statistical and systematic uncertainties have been combined),
although slightly higher, agrees with this value within the
uncertainty and indicates that Type Ia rates do not change
dramatically out to $z\sim$0.4.  Note, however that correcting for
host galaxy extinction and inclination may change this conclusion.

Theoretical estimates of Type Ia SNe rates have been derived from
stellar and galaxy evolution models. Calculations were done mostly for
elliptical galaxy type.  Recent calculations, based on evolutionary
models of elliptical galaxies, predict rates of $\sim 0.1 h^2$ SNu
(Ferrini \& Poggianti 1993). Assuming a factor of $\sim 2$ higher rate
in non-elliptical galaxies compared to ellipticals (Capellaro et
al. 1993b) and a mix of galaxy types as above, we convert this to an
overall rate of Type Ia SNe at $z \sim 0.4$ in all galaxy types, and
derive a value of $\sim 0.37 h^2$ SNu.  Our total uncertainty of
$^{+0.68}_{-0.49}$ in the measurement presented in this paper does not
allow any firm conclusion but our observed rate seems to lie above
this theoretical prediction. There may be an increase of Type Ia rate
with redshift.  Ruiz-Lapuente, Burkert \& Canal (1995) predict
significant increase in rate for redshifts between 0.4 and 0.8
depending on the specific model they consider.  In the near future,
our ongoing high-$z$ SN search and others should provide enough data
to constrain the theoretical calculations.

This work was supported in part by the
National Science Foundation (ADT-88909616) and the U.S. Dept. of
Energy (DE-AC03-76SF000098).  We thank the La Palma staff \& observers
for carrying out service observations. We also thank Simon Lilly and
Caryl Gronwall \& David Koo for providing their galaxy counts prior to
publication, and Richard Kron for useful discussions. I.M. Hook acknowledges
a NATO fellowship. R. Pain thanks Gerard Fontaine of CNRS-IN2P3 and Bernard
Sadoulet of CfPA, Berkeley for supporting this work.

\end{document}